\documentclass[smallextended]{svjour3}

\usepackage{graphicx, epsfig, amssymb} %include figure files
\usepackage{amsmath, amsfonts}
\usepackage{bm} %include bold math: \bm{} creates bold letters in math mode
\usepackage{color}
\usepackage[usenames]{xcolor}
\usepackage{hyperref}
\usepackage{siunitx}
\usepackage{cite}
\hypersetup{linktocpage,colorlinks=true,urlcolor=blue,linkcolor=blue,citecolor=blue}

\newcommand*{\diag}{\operatorname{diag}}
\newcommand{\diff}[1]{\text{d}#1}
\newcommand{\Diff}[1]{\text{D}#1}

\begin{document}

%%%%%%%
 
\title{Static and rotating black strings in dynamical Chern--Simons modified gravity}

%%%%%%%

\author{Adolfo Cisterna \and Crist\'obal Corral \and Sim\'on del Pino}

\institute{Adolfo Cisterna \at Universidad Central de Chile, Vicerrector\'ia acad\'emica, Toesca 1783, Santiago, Chile. \\
\email{adolfo.cisterna@ucentral.cl} \and 
Crist\'obal Corral \at Instituto de Ciencias Nucleares, Universidad Nacional Aut\'onoma de M\'exico, \\ Apartado Postal 70-543, Ciudad de M\'exico, 04510, M\'exico. \at Universidad Andr\'es Bello, Departamento de Ciencias F\'isicas, \\ Facultad de Ciencias Exactas, Sazi\'e 2212, Santiago, 8370136, Chile. \\ \email{cristobal.corral@correo.nucleares.unam.mx} \and 
Sim\'on del Pino \at Instituto de F\'isica, Pontificia Universidad Cat\'olica de Valpara\'iso, Casilla 4059, Valpara\'iso, Chile. \\ \email{simon.delpino.m@mail.pucv.cl}}

\maketitle

\begin{abstract}

Four-dimensional homogeneous static and rotating black strings in dynamical Chern--Simons modified gravity, with and without torsion, are presented. Each solution is supported by a scalar field that depends linearly on the coordinate that span the string. The solutions are locally ${\rm AdS}_3\times\mathbb{R}$ and they represent the continuation of the Ba\~nados--Teitelboim--Zanelli black hole. Moreover, they belong to the so-called Chern--Simons sector of the space of solutions of the theory, since the Cotton tensor contributes nontrivially to the field equations. The case with nonvanishing torsion is studied within the first-order formalism of gravity, and it considers nonminimal couplings of the scalar fields to three topological invariants: Nieh--Yan, Pontryagin and Gauss--Bonnet terms, which are studied separately. These nonminimal couplings generate torsion in vacuum, in contrast to Einstein--Cartan theory. In all cases, torsion contributes to an effective cosmological constant that, in particular cases, can be set to zero by a proper choice of the parameters. 
\end{abstract}
%%%%%%

\pagebreak

\section{Introduction}

Chern--Simons modified gravity (CSMG) is a well-known effective extension of general relativity (GR) that considers nonminimal coupling between a gravitational scalar degree of freedom and the topological Pontryagin density in four dimensions~\cite{Jackiw:2003pm}. This theory is motivated by anomaly cancelation in curved spacetimes, string theory and particle physics~\cite{Alexander:2009tp}. Its nonminimal coupling might explain flat galaxy rotation curves without introducing dark matter~\cite{Konno:2008np} and future gravitational wave detections might be sensitive to such a modification through frame dragging, gyroscopic precession, and amplitude birefringence in propagation of gravitational waves~\cite{Alexander:2007zg,Alexander:2007kv,Alexander:2017jmt}. Although it has been stablished that GW170817 falsifies some classes of modified gravity theories which dispense with the need of dark matter~\cite{Boran:2017rdn}, recently, it has been shown that CSMG is compatible with such experimental data~\cite{Nishizawa:2018srh}. When the scalar field is nondynamical, all spherically symmetric solutions of GR are solutions of CSMG, since the Pontryagin density and its associated Cotton tensor vanish identically,\footnote{In this context, the name Cotton tensor has been used to denote the contribution of the scalar-Pontryagin coupling to the Einstein's field equations, in analogy to Chern--Simons theory in three dimensions~\cite{Jackiw:2003pm} (see also~\cite{Garcia-Diaz:2017cpv}).} in contrast to axially-symmetric configurations~\cite{Grumiller:2007rv,Konno:2007ze,Yunes:2009hc}. In the dynamical case, on the other hand, CSMG should be considered as an effective theory, since there seems to be evidence that its Cauchy initial-value problem is ill-posed~\cite{Delsate:2014hba}. This, of course, is not an issue when stationary configurations are considered and it may be overcome within the first-order formulation of CSMG with nontrivial torsion.

In the realm of the first-order formalism of gravity, where the vierbein and Lorentz connection are regarded as independent fields, Chern--Simons modified gravity has been studied in different scenarios. In this framework, the nonminimal couplings of gravitational scalar degrees of freedom to topological densities generate torsion in vacuum, in contrast to Einstein--Cartan theory. Torsional invariants such as the Nieh--Yan, Pontryagin, and Euler densities, have been discussed in~\cite{Hehl:1994ue}. The scalar-Pontryagin coupling exhibits modifications to the standard four-fermion interaction appearing in Einstein--Cartan theory, whose physical consequences have been discussed in~\cite{Alexander:2008wi}. On the other hand, the Nieh--Yan\footnote{Although the name Nieh--Yan density is used throughout this work, it is worth mentioning that the pseudoscalar curvature in presence of torsion has been already discussed in Refs.~\cite{Hojman:1980kv,Nelson:1980ph} (see also~\cite{Mielke:2009zz}).} density~\cite{Nieh:1981ww} contributes to the fermionic axial anomaly in Riemann--Cartan spacetimes and it diverges once the regulator is removed~\cite{Chandia:1997hu}. This divergence has been debated in Refs.~\cite{Kreimer:1999yp,Chandia:1999az} and, later, a regularization procedure based on the scalar-Nieh--Yan coupling was formulated~\cite{Mercuri:2009zi,Ertem:2009ur}. When torsion is integrated out, this model resembles the scenarios of Refs.~\cite{Mielke:2006zp,Duncan:1992vz}, and it might offer a solution to the strong CP problem~\cite{Lattanzi:2009mg,Castillo-Felisola:2015ema,Karananas:2018nrj}. Finally, cosmological scenarios have been studied in the first-order formulation of scalar-Gauss--Bonnet gravity~\cite{Toloza:2013wi,Espiro:2014uda,Castillo-Felisola:2016kpe,Cid:2017wtf}, motivated by dimensional reductions of Lovelock gravity, which appears as low-energy corrections of string theory~\cite{Zwiebach:1985uq}. 

In this work it is shown that the spectrum of solutions of CSMG can be enlarged by the existence of black string configurations, with and without torsion. Black strings are, in principle, higher-dimensional asymptotically flat black hole solutions with an extended horizon of topology $S^2 \times R^n$, or $(S^1)^n$ when compactifying the extra dimensions~\cite{Obers:2008pj}. They are easily constructed by considering flat extra directions on the spacetime metric and they represent the most simple counterexample to the uniqueness theorems for higher dimensional GR~\cite{Carter:1971zc,Israel:1967za,Wald:1971iw}. Even more, they pave the way to construct asymptotically flat solutions with nonspherical topology, demonstrating that topological restrictions~\cite{Friedman:1993ty} lose their strength in higher dimensions. In spite of the evident simplicity involved in the addition of extra flat directions for constructing the black string, there are setups in which such a process is not evident. The most illustrative case is when the cosmological constant is included. It is direct to see that, if a $D=d+p$ dimensional spacetime is considered with $p$ flat directions, the field equations force the cosmological constant to vanish. This implies that there is no simple black string extension of the Schwarzschild--(A)dS black hole.\footnote{Nonhomogeneous AdS black strings have been constructed in Ref.~\cite{Chamblin:1999by} by considering warped spacetimes. This result was generalized for Lovelock theories possessing a unique constant curvature vacua~\cite{Kastor:2006vw} and for more general Lovelock theories by generalizing the concept of Einstein spaces~\cite{Kastor:2017knv}. AdS black strings and black rings have been constructed only numerically for nonhomogeneous geometries~\cite{Obers:2008pj,Horowitz:2012nnc}.} Similar obstructions are encountered for the black string extension of the Reissner--Nordstrom black hole. 

Here, the approach of~\cite{Cisterna:2017qrb} is generalized to show that exact four-dimensional black string vacuum solutions, with and without torsion, can be obtained in CSMG. In order to do so, the scalar fields are assumed to depend only on the extended coordinate that span the string, in contrast to Refs.~\cite{Ahmedov:2010rn,Ahmedov:2010fz}. This assumption allows us to find solutions to the field equations that belong to the so-called Chern--Simons sector of the space of solutions of the theory~\cite{Grumiller:2007rv}, since the nonminimal coupling between the scalar field and the Pontryagin density contributes nontrivially to the field equations. Then, the first-order formulation of CSMG is studied by considering the nonminimal couplings of the scalar fields to the Nieh--Yan, Pontryagin and Gauss--Bonnet terms, and black strings with nonvanishing torsion are found in each case.

The article is organized as follows: in Sec.~\ref{sec:rbs}, the method for constructing black strings in GR in arbitrary dimensions is reviewed, the Riemannian (torsion-free) dynamical CSMG theory is presented, and the Ba\~nados--Teitelboim--Zanelli (BTZ) black string solution is obtained. In Sec.~\ref{sec:focsmg}, we obtain the black string solutions with nontrivial torsion by taking separately the nonminimal couplings of scalar fields to Nieh--Yan, Pontryagin and Gauss--Bonnet terms. Conclusions and comments are presented in Sec.~\ref{sec:conclusions} and two Appendices have been included for details. Henceforth, greek and latin characters represent spacetime and Lorentz indices, respectively, and the metric signature is $(-,+,+,+)$. 

\section{Black strings in dynamical Chern--Simons modified gravity\label{sec:rbs}}

In this section, static and rotating  BTZ black strings in CSMG with a dynamical scalar field are presented. Configurations of this class have been studied in the nondynamical case in~\cite{Ahmedov:2010fz}, by assuming that the scalar field depends on the radial coordinate of the $3$-dimensional section of the spacetime. In such a case, the scalar field remains arbitrary and it acts as a Lagrange multiplier that imposes the Pontryagin density to vanish. Even though this constraint is trivially fulfilled by virtue of the isometry group, the radial dependence of the scalar field implies that the Cotton tensor does not contribute to the Einstein's field equations~\cite{Ahmedov:2010fz}. In the present work, unlike~\cite{Ahmedov:2010fz}, the scalar field is dynamical and it can be solved exactly from the field equations. These configurations belong to the so-called Chern--Simons sector of the space of solutions~\cite{Grumiller:2007rv} and they present the main goal of this section.

\subsection{Homogeneous AdS black strings in GR\label{sec:method}}

Before presenting the BTZ black strings in CSMG, we briefly review the procedure of~\cite{Cisterna:2017qrb} to obtain homogenous anti-de Sitter black strings in GR.

Let us consider a spacetime metric of dimension $D=d+p$, where $p$ flat extended coordinates have been added to a $d$-dimensional metric according to
\begin{align}\label{metricstructure}
\diff{s}^2&=\diff{\bar{s}}^2+\delta_{ij}\diff{z}^i \diff{z}^j,   
\end{align} 
where $\diff{\bar{s}}^2$ stands for the $d$-dimensional metric, $\delta_{ij}$ is $p$-dimensional Euclidean metric, and $i=1,\ldots,p$. In order to obtain homogenous configurations, it is demanded that $\diff{\bar{s}}^2$ does not depend on the extra coordinates $z^i$. No further conditions are required, nonetheless, for simplicity we consider 
\begin{equation}
\diff{\bar{s}}^2=-f(r)\diff{t}^2+  \frac{\diff{r}^2}{f(r)} + r^2\diff{\Omega}^2_{d-2,\gamma}, \label{ans2}
\end{equation}
where $\diff{}\Omega^2_{d-2,\gamma}$ represents a $(d-2)$-dimensional base manifold of constant curvature $\gamma=0,\pm1$, representing flat, spherical, and hyperbolic geometries respectively. 

In order to circumvent the unavoidable vanishing of the cosmological constant imposed by the field equations, the authors of~\cite{Cisterna:2017qrb} have dressed each of the $p$ extended coordinates $z^i$ with a minimally coupled scalar field $\varphi_i$, with $i=1...p$, depending only on those coordinates, i.e., $\varphi_i=\varphi_i(z^i)$. The field equations are then written as
\begin{align}\label{eomgr}
 R_{\mu\nu} - \frac{1}{2}g_{\mu\nu}R + \Lambda g_{\mu\nu} &= \kappa \tau^{(\varphi)}_{\mu\nu},\\
 \label{KG}
 \Box\varphi_i&=0.
\end{align}
The Klein--Gordon equation associated to each $\varphi_i$ is easily integrated providing scalar fields with a linear dependence on the $z^i$ coordinates. By using the remaining rotational symmetry, they can be cast into the form 
\begin{equation}
\varphi^i=\lambda z^i,     \label{sca}
\end{equation}
where $\lambda$ is an integration constant which is usually called the axionic charge. This kind of scalar fields have been used to construct planar hairy black holes that exhibit momentum relaxation in their dual representations. Due to the explicit breaking of translational symmetry, it is possible to obtain well behaved holographic conductivities in the dual field theory~\cite{Andrade:2013gsa}.

The Einstein field equations projected along the $d$-dimensional manifold and the flat coordinates take the form
\begin{align}
\bar{R}_{\bar{\mu}\bar{\nu}} - \frac{1}{2}\bar{g}_{\bar{\mu}\bar{\nu}}\bar{R} &= - \left(\Lambda+\frac{\kappa p\lambda^2}{2}\right)\bar{g}_{\bar{\mu}\bar{\nu}},\label{pro1}\\
\bar{R}&=2\Lambda-2\kappa\left(1-\frac{p}{2}\right)\lambda^2, \label{pro2}
\end{align}
respectively, where bar objects represent quantities defined on the $d$-dimensional section. Compatibility of the field equations~\eqref{pro1} and~\eqref{pro2} requires the following relation to hold
\begin{equation}
\lambda^2=-\frac{2\Lambda}{\kappa(d+p-2)}.  \label{cons1} 
\end{equation}
From here it is direct to see that, in absence of the scalar fields, the cosmological constant would have vanish. 

Replacing~\eqref{cons1} in~\eqref{pro1}, the metric function is found to be
\begin{equation}
f(r)= \gamma-\frac{M}{r^{d-3}}-\frac{2\Lambda r^2}{(d-1)(d+p-2)},
\end{equation}
where $M$ is an integration constant related to the mass. Observe that $\Lambda$ must be negative and that the AdS radius gets a modification given by the number of extra flat directions $p$. This result imply that black strings, which are originally thought to be higher dimensional objects, also exist in four dimensions. In fact, the BTZ black string metric reads
\begin{equation}
\diff{s}^2= -\left(-M-\frac{\Lambda r^2}{2}\right)\diff{t}^2+\frac{\diff{r}^2}{\left(-M-\frac{\Lambda r^2}{2}\right)}+r^2\diff{\phi}^2+\diff{z}^2,
\end{equation}
which is supported by a single scalar field $\varphi=\lambda z$, with $\lambda^2=-\Lambda/\kappa$. In what follows, these ingredients are used to construct homogenous BTZ black strings in dynamical CSMG by considering $d=3$ and $p=1$.

\subsection{Chern--Simons modified gravity\label{sec:rcsmg}}

Chern--Simons modified gravity considers two independent gravitational fields: the metric $g_{\mu\nu}$ and the scalar field $\varphi$. The action principle for the dynamical CSMG is given by~\cite{Alexander:2009tp}\footnote{In Appendix~\ref{sec:relation} the theory is presented in the language of differential forms to connect with the Sec.~\ref{sec:focsmg} of the manuscript.}
\begin{align}\notag
 S\left[g_{\mu\nu},\varphi \right] &= \frac{1}{2\kappa}\int\diff{}^4x\sqrt{-g}\left(R - 2\Lambda + \frac{\alpha}{4}\varphi\,^*RR\right) \\
 \label{riemaction}
 &\quad- \frac{1}{2}\int\diff{}^4x\sqrt{-g}\;g^{\mu\nu}\nabla_\mu\varphi\nabla_\nu \varphi ,
\end{align}
where $\kappa=8\pi G_N$ is the gravitational constant, $\alpha$ is a dimensionful coupling constant, and the Pontryagin term is 
\begin{align}
 ^*RR \equiv \frac{1}{2}\epsilon^{\gamma\delta\tau\sigma} R^{\mu}{}_{\nu\gamma\delta} R^{\nu}{}_{\mu\tau\sigma},
\end{align}
with $\epsilon^{\mu\nu\lambda\rho}$ being the Levi-Civita tensor. The field equations for this theory are obtained by performing stationary variations of the action~\eqref{riemaction} with respect to the metric and the scalar field, respectively giving
\begin{align}\label{eomg}
 R_{\mu\nu} - \frac{1}{2}g_{\mu\nu}R + \Lambda g_{\mu\nu} + \alpha\, C_{\mu\nu} &= \kappa \tau^{(\varphi)}_{\mu\nu},\\
 \label{eomp}
 \Box\varphi + \frac{\alpha}{8\kappa}\, ^*RR &= 0,
\end{align}
where
\begin{align}
 \tau^{(\varphi)}_{\mu\nu} &= \nabla_\mu\varphi\nabla_\nu\varphi - \frac{1}{2}g_{\mu\nu}\nabla_\lambda\varphi\nabla^\lambda\varphi,\\
 C^{\mu\nu} &= \nabla_\rho\varphi\,\epsilon^{\rho\sigma\lambda(\mu}\nabla_\lambda R^{\nu)}{}_\sigma + \nabla_\rho\nabla_\sigma\varphi\, ^*R^{\sigma(\mu\nu)\rho}, \label{coton}
\end{align}
and $^*R^{\mu\nu\lambda\rho} = \tfrac{1}{2}\epsilon^{\lambda\rho\sigma\tau}R^{\mu\nu}{}_{\sigma\tau}$. Notice that the contribution of the Cotton tensor $C_{\mu\nu}$ in the Einstein's field equations involves covariant derivatives of the Riemann tensor, giving, in general, third order field equations for the metric. Importantly, since the field equations of CSMG involve only derivatives of the scalar field, they are invariant under the shift $\delta\phi = \phi_0$, where $\phi_0$ is a constant, while the metric remains invariant. This is a key feature of CSGM that will be useful to construct black string configurations supported by scalar fields with a non-trivial linear dependence along the string extended coordinate.

\subsection{Static BTZ black string}
In order to look for static BTZ black string solutions to the field equations~\eqref{eomg} and~\eqref{eomp}, the following metric ansatz is considered
\begin{align}\label{metricansatz}
\diff{s}^2 &= -f(r)\diff{t}^2 + \frac{\diff{r}^2}{f(r)} + r^2\diff{\phi}^2 + \diff{z}^2.
\end{align}
Importantly, the Pontryagin density vanishes identically for this metric ansatz, i.e., $^*RR=0$. The latter implies a free Klein--Gordon equation for $\varphi$, that can be easily integrated as $\varphi=\lambda z$.

Due to the staticity of the metric and the form of the scalar field, the Cotton tensor~\eqref{coton} has only one nonvanishing component, that is
\begin{equation}
C_{t\phi}=\frac{\lambda r f}{4}f''',  \label{co3}
\end{equation}
where prime denote derivative with respect to the coordinate $r$ and all diagonal components of the Cotton tensor vanish. This implies that the rest of the field equations are given by~\eqref{pro1} and~\eqref{pro2} for $d=3$ and $p=1$, which will be compatible only if 
\begin{equation}
\lambda^2=-\Lambda/\kappa.  \label{consis}
\end{equation}
Then, by solving the non-diagonal components of the field equations, namely $C_{t\phi}=0$, and using~\eqref{consis}, the static BTZ black string solution that solves the full system of field equations is found to be
\begin{equation}\label{solstatic}
\diff{s}^2=-\left(\frac{r^2}{l^2}-M\right)\diff{}t^2+\frac{\diff{}r^2}{\left(\frac{r^2}{l^2}-M\right)}+r^2\diff{}\phi^2+\diff{}z^2,
\end{equation}
where $l^{-2}=-\Lambda/2$. This is the first example of a BTZ black string for dynamical CSMG with a nontrivial contribution of the Cotton tensor and it exists due to the linear dependence of the scalar field on the extended coordinate as it can be seen from~\eqref{co3}. 
This solution is locally $\text{AdS}_3\times\mathbb{R}$, and it represents the cylindrical extension of the BTZ black hole~\cite{Banados:1992wn}. It must be noticed that for a scalar field depending on the radial coordinate only, i.e. $\varphi=\varphi(r)$, the Cotton tensor vanishes identically, and the scalar field cannot be determined from the field equations~\cite{Ahmedov:2010fz}.

\subsection{Rotating BTZ black string}
It is observed that the machinery described in the previous section can be also used to construct rotating BTZ black strings in dynamical CSMG. To accomplish this, the following metric ansatz is considered
\begin{align}
\diff{s}^2 &= -f(r)\diff{t}^2 + \frac{\diff{r}^2}{f(r)} + r^2(N(r)\diff{t}+\diff{\phi})^2 + \diff{z}^2, \label{rota}
\end{align}
where the addition of an extended flat coordinate has been applied to a rotating $3$-dimensional metric ansatz. Since the Pontryagin density vanishes identically for this ansatz, i.e. $^*RR=0$, and under the assumption that the scalar field depends only on the extended coordinate, the Klein--Gordon equation reduces to a free scalar field that can be solved as $\varphi=\lambda z$. Nevertheless, rotation has a highly nontrivial effect in the rest of the field equations of dynamical CSMG, since the Cotton tensor contributes to all diagonal field equations for the metric along the $3$-dimensional black hole spacetime. 

In order to obtain rotating BTZ black strings for~\eqref{eomg}, it is noticed that the nontrivial components of the Cotton are
\begin{align}
\notag
 C_{tt} &= \frac{\lambda f}{4r}\bigg(f'' N' r^2 - f' N''r^2 - 4f'N' r - 2N'''fr^2 - 10N''fr
 \\
 &\quad - 2(N')^3r^4 - 6N'f' \bigg),\\
 \notag
 C_{rr} &= \frac{\lambda}{4rf}\bigg(-f''N'r^2 + f'N''r^2+4f'N'r - 2N''fr \\
 &\quad + 2(N')^3r^4 - 6N'f\bigg),\\
 C_{t\phi} &= \frac{\lambda rf}{4}\bigg(f''' - 6N'' N'r^2 - 12(N')^2 r\bigg),\\
 C_{\phi\phi} &= \frac{\lambda r^2}{2}\bigg(f''N'r - f'N''r - 4f'N' - N'''fr - 4N''f - 2(N')^3r^3 \bigg),
\end{align}
where, as before, prime denotes derivative with respect to the coordinate $r$. In order to integrate the field equations, first, the condition $C_{\mu\nu} = 0$ is solved and one finds
\begin{align}
\label{funcion1}
N(r)&=j_1 +\frac{j_2}{r}+\frac{J}{r^2},\\    
f(r)&=  f_1r^2 +  f_2\left(j_2r + J\right) + \frac{\left(j_2r + J\right)^2}{r^2} ,\label{funcion2} 
\end{align}
where $j_1$, $j_2$, $J$, $f_1$ and $f_2$ are integration constants. Then, the field equations are given again by~\eqref{pro1} and~\eqref{pro2} with $d=3$ and $p=1$. Replacing $\varphi = \lambda z$ with the compatibility condition~\eqref{consis} along with~\eqref{funcion1} and~\eqref{funcion2} in Eq.~\eqref{eomg}, we obtain that $j_2=0$ and $f_1=-\Lambda/2$; moreover, the integration constant $j_1$ can be set to zero without loss of generality~\cite{Banados:1992gq1}. Thus, the rotating BTZ black string is given by Eq.~\eqref{rota} with
\begin{align}
f(r) = -M+\frac{r^2}{l^2} +\frac{J^2}{r^2} \;\;\; \mbox{and} \;\;\; N(r) = \frac{J}{r^2}, \label{solrota}
\end{align}
where we have defined $M\equiv-Jf_2$ and $J$ such that they represent the integration constants related to mass and the angular momentum, respectively, and $l^{-2}=-\Lambda/2$. This solution represents a full rotating exact black string solution to dynamical CSMG. Notice that there is no continuous limit to the static case by taking $J\to0$, since the mass vanishes and it yields to a naked singularity. As well as in the static case, rotating BTZ black strings were previously found only for the non-dynamical case, where $\varphi=\varphi(r)$ acts merely as a Lagrangian multiplier and it remains undetermined from the field equations~\cite{Ahmedov:2010fz}. 

The horizon structure of the rotating BTZ black string is governed by $f(r_\pm)=0$, with
\begin{align}
 r_\pm = \ell\left[\frac{M \pm \sqrt{M^2 - 4J^2/\ell^2}}{2}\right]^{\frac{1}{2}}.
\end{align}
For a horizon to exist, the condition $M^2 - 4J^2/\ell^2\geq0$ must be met, with $J\neq0$. The curvature invariants constructed out of~\eqref{metricansatz} remain constants, however, a singularity at $r=0$ arises from the identification of points of anti-de Sitter space by a discrete subgroup of $SO(2,2)$~\cite{Banados:1992gq1}.

This solution is also supported if the scalar-Gauss--Bonnet coupling is added to the CSMG action~\eqref{riemaction}. Since the Gauss--Bonnet term
\begin{align}
 \mathcal{G} = R^2 - 4 R_{\mu\nu}R^{\mu\nu} + R_{\mu\nu\lambda\rho}R^{\mu\nu\lambda\rho},
\end{align}
vanishes identically for the metric ansatz~\eqref{rota}, the field equation for the scalar field reduces to the free Klein--Gordon equation, which can be solved as \mbox{$\varphi=\lambda z$}. With this solution, the contribution of the scalar-Gauss--Bonnet coupling to the Einstein equations vanishes and the system reduces to~\eqref{pro1} and~\eqref{pro2} with $d=3$ and $p=1$, whose solution is given in~\eqref{solrota}. The addition of the scalar-Gauss--Bonnet coupling, however, is nontrivial in the case with nonvanishing torsion as it is shown in the next section.

\section{Black strings in torsional Chern--Simons modified gravity\label{sec:focsmg}}

In order to deal with spacetimes with torsion, we will work within the first-order formalism of gravity that considers two independent gravitational fields: the vierbein $1$-form \mbox{$e^a = e^{a}{}_\mu \diff{x^\mu}$}, that encodes the spacetime metric through \mbox{$g_{\mu\nu} = \eta_{ab}e^{a}{}_{\mu} e^{b}{}_\nu$}, where $\eta_{ab}=\diag(-,+,+,+)$; and the Lorentz connection $1$-form \mbox{$\omega^{ab}= \omega^{ab}{}_\mu \diff{x^\mu}$}, encoding its affine structure. These fields transform as $1$-forms under diffeomorphisms and as a vector and gauge connection under local Lorentz transformations, respectively.  The Lorentz curvature and torsion $2$-forms are defined by the Cartan's structure equations
\begin{align}
\label{curvature}
 R^{ab} &= \diff{\omega^{ab}} + \omega^{a}{}_c\wedge\omega^{cb},\\
\label{torsion}
 T^a &= \diff{e^a} + \omega^{a}{}_b\wedge e^b \equiv \Diff{e^a},
\end{align}
where $\diff{}$ is the exterior derivative, $\wedge$ is the wedge product of differential forms, and $\Diff{}$ is the exterior Lorentz-covariant derivative. Curvature and torsion satisfy the Bianchi identities $\Diff{R^{ab}} = 0$ and $\Diff{T^a} = R^{a}{}_b\wedge e^b$. 

\subsection{First-order Chern--Simons modified gravity}

The first-order formulation of CSMG is considered by including extra scalar fields nonminimally coupled to the Nieh--Yan, Pontryagin, and Gauss--Bonnet terms. It is described by the action principle
\begin{align}\notag
 S[e^a,\omega^{ab},\varphi_i] &= \frac{1}{4\kappa}\int\left[\epsilon_{abcd}\left(R^{ab} - \frac{\Lambda}{6}e^a\wedge e^b\right)\wedge e^c\wedge e^d + \sum_{i=1}^3\alpha_i\varphi_i\mathcal{I}_i  \right] \\ 
 \label{focsmga}
 &\quad - \frac{1}{2}\int\sum_{i=1}^3 \diff{\varphi_i}\wedge\star\diff{\varphi_i},
\end{align}
where $\kappa=8\pi G_N$, $\Lambda$ is the cosmological constant, $\star$ is the Hodge dual, and
\begin{subequations}\label{invariants}
\begin{align}
 \mathcal{I}_1 &= T_a\wedge T^a - R_{ab}\wedge e^a\wedge e^b = \diff{}\left(e_a\wedge T^a \right),\\
 \mathcal{I}_2 &= R^{a}{}_b\wedge R^{b}{}_a = \diff{}\left[\omega^{a}{}_b\wedge\left(R^{b}{}_a - \frac{1}{3}\omega^{b}{}_c\wedge\omega^{c}{}_a \right) \right],\\
 \mathcal{I}_3 &= \epsilon_{abcd}R^{ab}\wedge R^{cd} = \diff{}\left[\epsilon_{abcd}\,\omega^{ab}\wedge\left(R^{cd} - \frac{1}{3}\omega^{c}{}_f\wedge\omega^{fd} \right) \right],
\end{align}
\end{subequations}
denote the Nieh--Yan, Pontryagin, and Gauss--Bonnet densities, respectively. Here, the index $i$ denote different nonminimal couplings to topological invariants rather than extended coordinates as in Subsec.~\ref{sec:method}, and $\alpha_i$ are dimensionful coupling constants. These models have been considered in cosmological scenarios presenting interesting phenomenology~\cite{Toloza:2013wi,Espiro:2014uda,Cid:2017wtf}.

The field equations are obtained by performing stationary variations of~\eqref{focsmga} with respect to the vierbein, Lorentz connection, and scalar fields giving
\begin{subequations}\label{eomfocsmg}
\begin{align}\label{eome}
 \mathcal{E}_a &\equiv \epsilon_{abcd}\left(R^{bc} - \frac{\Lambda}{3} e^b\wedge e^c\right)\wedge e^d + \alpha_1\diff{\varphi_1}\wedge T_a - 2\kappa\sum_{i=1}^3\tau_a^{(\phi_i)} =0,\\
 \notag
 \mathcal{E}_{ab} &\equiv \epsilon_{abcd}T^c\wedge e^d - \frac{\alpha_1}{2}\diff{\varphi_1}\wedge e_a\wedge e_b - \alpha_2\diff{\varphi_2}\wedge R_{ab} \\
 \label{eomw}
 &\quad + \alpha_3\epsilon_{abcd}\diff{\varphi_3}\wedge R^{cd} =0,\\
 \label{eoms}
 \mathcal{E}^{(\varphi_i)} &\equiv \diff{\star}\diff{\varphi_i} + \frac{\alpha_i}{4\kappa}\mathcal{I}_i = 0,
\end{align}
 \end{subequations}
respectively, where no sum over $i$ is assumed unless stated otherwise. From Eq.~\eqref{eomw} it can be seen that torsion is sourced by the exterior derivative of the scalar fields.  The energy-momentum $3$-form of each scalar field is defined as
\begin{align}\label{tauscal}
 \tau_a^{(\varphi_i)} &= -\frac{1}{2}\bigg(\diff{\varphi_i}\wedge\star\left(\diff{\varphi_i}\wedge e_a\right) + \left(i_a\diff{\varphi_i}\right)\star\diff{\varphi_i}\bigg).
\end{align}

Diffeomorphism invariance imply the on-shell conservation law for each scalar field as
\begin{align}
 \Diff{\tau_a^{(\varphi_i)}} &= \left(i_a T^b\right)\wedge \tau_b^{(\varphi_i)} + \frac{\alpha_i}{4\kappa}i_a\diff{\varphi_i}\,\mathcal{I}_i,
\end{align}
where $i=1,2,3$. Invariance under local Lorentz transformations, on the other hand, imply a condition that is trivially satisfied for the energy-momentum $3$-form of the scalar fields. Moreover, the action~\eqref{focsmga} is quasi-invariant under the global shift symmetry $\delta\varphi_i = \varphi_0^{(i)}$, where $\varphi_0^{(i)}$ are constants. The Noether current associated to this symmetry, $J_i=\star\diff{\varphi_i} + \tfrac{\alpha_i}{4\kappa}C_i$, is conserved on shell by virtue of the field equation~\eqref{eoms}, where the $C_i$'s have been defined as $\mathcal{I}_i = \diff{C_i}$ according to Eq.~\eqref{invariants}.

\subsection{BTZ black strings with nonvanishing torsion\label{sec:btzbswt}}

The vierbein compatible with the metric structure~\eqref{metricstructure} can be written as
\begin{align}\label{vierbeinansatz}
e^a &= \begin{cases}
       e^{\bar{a}} = \bar{e}^{\bar{a}}(\bar{x}),\\
       e^3 = \diff{z},
       \end{cases}
\end{align}
where barred quantities denote projection on the $3$-dimensional spacetime.\footnote{For instance, $\bar{a}=0,1,2$ denote $3$-dimensional Lorentz indices, $\bar{x}$ their local coordinates, and $\bar{e}^{\bar{a}}$ is recognized as the dreibein.} The Lorentz connection, on the other hand, contains torsional degrees of freedom beyond the metric ones, and it accepts a decomposition compatible with~\eqref{vierbeinansatz} given by
\begin{align}\label{spinansatz}
\omega^{ab} &= \begin{cases}
               \omega^{\bar{a}\bar{b}} = \bar{\omega}^{\bar{a}\bar{b}}+\alpha^{\bar{a}\bar{b}}e^3,\\
	       \omega^{\bar{a}3} = \beta^{\bar{a}}+\gamma^{\bar{a}}e^3,
              \end{cases}
\end{align} 
where the fields $\omega^{\bar{a}\bar{b}}$, $\alpha^{\bar{a}\bar{b}}$, $\beta^{\bar{a}}$ and $\gamma^{\bar{a}}$ depend only on $\{\bar{x}\}$. The piece $\bar{\omega}^{\bar{a}\bar{b}}$ is recognized as the Lorentz connection of the $3$-dimensional spacetime, $\alpha^{\bar{a}\bar{b}}=-\alpha^{\bar{b}\bar{a}}$ and $\gamma^{\bar{a}}$ are Lorentz-valued $0$-forms, while $\beta^{\bar{a}}=\beta^{\bar{a}}{}_{\bar{b}}e^{\bar{b}}$ is a Lorentz-valued $1$-form. It is worth noticing that, even though the topological invariants constructed out of the Levi-Civita connection vanish by virtue of the isometries of~\eqref{metricansatz}, this is not the case when a torsionful connection compatible with such isometries is considered. 

The curvature and torsion, as well as the equations of motion, are decomposed in terms of the transverse section and the extended direction. The details of such a decomposition is given in Appendix~\ref{decomp} and, from hereon, we adopt such a notation. The distinctive parts of the field equations~\eqref{eomfocsmg}, namely, $\mathcal{A}_{\bar{a}}$, $\mathcal{B}_{\bar{a}}$, $\mathcal{C}$, $\mathcal{D}$, $\mathcal{W}_{\bar{a}\bar{b}}$, $\mathcal{X}_{\bar{a}\bar{b}}$, $\mathcal{Y}_{\bar{a}}$, and $\mathcal{Z}_{\bar{a}}$, must vanish independently on shell. 

In this work, the following metric ansatz for the $3$-dimensional manifold is considered
\begin{align}\label{3ansatz}
\bar{e}^0 &= f(r)\diff{t}, & \bar{e}^1 &= h(r)\diff{r}, & \bar{e}^2 &= r\left(N(r)\diff{t}+\diff{\phi}\right).
\end{align}
When the isometry group of~\eqref{3ansatz} is demanded to the torsional degrees of freedom, it is found that 24 nonvanishing independent components of $\omega^{ab}$ appear through the fields $\bar{\omega}^{\bar{a}\bar{b}}$, $\alpha^{\bar{a}\bar{b}}$, $\beta^{\bar{a}}$, and $\gamma^{\bar{a}}$. 

We are interested in solutions possessing scalar fields that depend only on the extended flat direction, namely, $\varphi_i = \varphi_i(z)$. According to the notation of Appendix~\ref{decomp}, this condition implies that $\bar{\tau}^{(i)}_{\bar{a}} = 0$, $\nu^{(i)} = 0$, and $\chi^{(i)}_{\bar{a}}\wedge e^{\bar{a}} = 3\mu^{(i)}$. From $\mathcal{W}_{\bar{a}\bar{b}} = 0$ and $\mathcal{Y}_{\bar{a}} = 0$ one has
\begin{align}
\beta_{\bar{a}} \wedge e^{\bar{a}} &= 0, \\
\epsilon_{\bar{a}\bar{b}\bar{c}} \bar{T}^{\bar{b}}\wedge e^{\bar{c}} &= 0,
\end{align}
where $\epsilon_{\bar{a}\bar{b}\bar{c}} \equiv \epsilon_{abc3}$. This, in turn, gives $\epsilon_{\bar{a}\bar{b}\bar{c}} L^{\bar{b}}\wedge e^{\bar{c}} = 0$ and $\epsilon_{\bar{a}\bar{b}\bar{c}}N^{\bar{a}\bar{b}}\wedge e^{\bar{c}} = 0$, as a consequence of $\mathcal{A}_{\bar{a}} = 0$ and $\mathcal{D}=0$. These equations can be used to integrate $\beta^{\bar{a}}$ as
\begin{align}\label{betasol}
 \beta^{\bar{a}} &= \beta_0 \bar{e}^{\bar{a}},
\end{align}
where $\beta_0$ is an integration constant. Additionally, the fact that $\mathcal{X}_{\bar{a}\bar{b}}\wedge e^{\bar{b}} = 0$ yields to
\begin{align}\label{eqgam}
 \epsilon_{\bar{a}\bar{b}\bar{c}}\gamma_{\bar{d}} e^{\bar{d}}\wedge e^{\bar{c}}\wedge e^{\bar{b}} = \alpha_2 \partial_z \varphi_2 \bar{R}_{\bar{a}\bar{b}}\wedge e^{\bar{b}}.
\end{align}
The Bianchi identity $\bar{\Diff{}}\bar{T}^{\bar{a}} = \bar{R}^{\bar{a}}{}_{\bar{b}}\wedge e^{\bar{b}}$ implies that Eq.~\eqref{eqgam} can be solved as
\begin{align}\label{gamsol}
 \gamma_{\bar{a}} = \frac{\alpha_2}{2}\partial_z \varphi_2 \bar{\star}\bar{\Diff{}}\bar{T}_{\bar{a}},
\end{align}
where $\bar{\star}$ is the Hodge dual with respect to $\bar{e}^{\bar{a}}$. Following a similar procedure, the equation $\mathcal{Z}_{\bar{a}}=0$ can be used to find
\begin{align}\label{alphasol}
 \alpha_{\bar{a}\bar{b}} &= -\alpha_3\partial_z \varphi_3 \epsilon_{\bar{a}\bar{b}\bar{c}}\bar{\star}\bar{\Diff{}}\bar{T}^{\bar{c}}.
\end{align}

In what follows, three cases are studied separately: (i)~scalar-Nieh--Yan coupling, obtained when $\alpha_2=0$, $\alpha_3=0$, $\varphi_2 =0$, and $\varphi_3= 0 $, (ii)~scalar-Pontryagin coupling, obtained when $\alpha_1=0$, $\alpha_3=0$, $\varphi_1 =0$, and $\varphi_3= 0 $, and (iii)~scalar-Gauss--Bonnet coupling, obtained when $\alpha_1=0$, $\alpha_2=0$, $\varphi_1=0$, and $\varphi_2=0$. Since our interest is to find locally $\mbox{AdS}_3\times\mathbb{R}$ black string solutions, we restrict ourselves to locally $\mbox{AdS}_3$ Riemann--Cartan spacetimes on the $3$-dimensional section, i.e.,
\begin{align}\label{rcads}
 \bar{R}^{\bar{a}\bar{b}}_{(i)} &= -\frac{1}{\ell_i^2}\bar{e}^{\bar{a}}\wedge\bar{e}^{\bar{b}},
\end{align}
where $\ell_i$'s denote the non-Riemannian AdS curvature radii for each different nonminimal coupling denoted by the subscript $i$, with $i=1,2,3$. For this class of spacetimes, the Bianchi identity implies that $\bar{\Diff{}}\bar{T}^{\bar{a}} = 0$. Thus, from Eqs.~\eqref{gamsol} and~\eqref{alphasol} one concludes that $\gamma_{\bar{a}}$ and $\alpha_{\bar{a}\bar{b}}$ vanish. The condition $\bar{\Diff{}}\bar{T}^{\bar{a}} = 0$ can be integrated, giving~\cite{Alvarez:2014uda}
\begin{align}\label{bartsol}
 \bar{T}^{\bar{a}} = \frac{1}{2}t_0\epsilon^{\bar{a}}{}_{\bar{b}\bar{c}} \bar{e}^{\bar{b}}\wedge\bar{e}^{\bar{c}},
\end{align}
where $t_0$ is an integration constant. Importantly, the Nieh--Yan, Pontryagin, and Gauss--Bonnet densities vanish for this class of spacetimes. Therefore, the Klein--Gordon equation can be solved in each case as
\begin{align}
 \varphi_i &= \lambda_i z,
\end{align}
where $\lambda_i$'s are integration constants associated to each scalar field.

The three cases are solved by~\eqref{vierbeinansatz}, with a $3$-dimensional section given by~\eqref{3ansatz} with 
\begin{align}\label{solBTZbs}
 f^2 = h^{-2} = -M + \frac{J^2}{r^2} + \frac{r^2}{\tilde{\ell}_i^2} \;\;\;\;\; \mbox{and} \;\;\;\;\; N = \frac{J}{r^2},
\end{align}
with $M$ and $J$ being integration constants related to the mass and angular momentum, while the three different effective Riemannian AdS curvature radii are denoted by $\tilde{\ell}_i$, with $i=1,2,3$. Moreover, in order to solve the field equations, we find that the integration constants $\beta_0$, $t_0$, and $\lambda_i$'s must be fixed in terms of the theory's parameters. These solutions represent the cilindrical extension of the BTZ black hole and they are summarized in Table~\ref{table:1}.
\begin{table}[h!]
\scalebox{0.9}{
\begin{tabular}{|c||c|c|c|}
 \hline
  & (i)~Pseudoscalar-Nieh--Yan & (ii)~Pseudoscalar-Pontryagin & (iii)~Scalar-Gauss--Bonnet \\
 \hline \hline
 $\lambda_i$ & $\lambda_1^2 = -\frac{8\Lambda}{3\alpha_1^2 + 8\kappa}$ & $\lambda_2^2 = -\frac{\Lambda}{\kappa}$ & $\lambda_3^2 = -\frac{\Lambda}{\kappa}$ \\ 
 $\beta_0$ & $0$ & $0$ & $\frac{\alpha_3\lambda_3\Lambda}{2}$ \\
 $t_0$ & $-\frac{\alpha_1\lambda_1}{2}$ & $-\frac{\alpha_2\lambda_2\Lambda}{2}$ & 0 \\
 $\frac{1}{\tilde{\ell}_i^2}$ & $-\frac{\Lambda}{2}$ & $-\frac{\Lambda}{2}\left(1 + \frac{\alpha_2^2\Lambda^2}{8\kappa} \right)$ & $-\frac{\Lambda}{2}\left(1-\frac{\alpha_3^2 \Lambda^2}{2\kappa} \right)$ \\
 $\frac{1}{\ell_i^2}$ & $-\frac{\Lambda\left(\alpha_1^2 + 4\kappa\right)}{3\alpha_1^2 + 8\kappa}$ & $- \frac{\Lambda}{2}$ & $-\frac{\Lambda}{2}$ \\
 \hline
 \end{tabular}}
 \caption{The BTZ black strings~\eqref{solBTZbs} solve the field equations~\eqref{eomfocsmg}, provided the torsional configurations~\eqref{betasol} and~\eqref{bartsol} with the integration constants $\beta_0$, $t_0$ and $\lambda_i$ fixed as shown in this table. The (non-)Riemannian AdS curvature radii ($\ell_i$) $\tilde{\ell}_i$ are presented for each case separately.}
 \label{table:1}
 \end{table}

The appearance of a modified Riemannian AdS curvature radius stem from the fact that $\beta^{\bar{a}}$, $\bar{T}^{\bar{a}}$, and $\varphi_i$ contribute to an effective cosmological constant, as it can be seen from equation of motion $\mathcal{B}_{\bar{a}}=0$. The contribution of torsion to the cosmological constant in $3$-dimensions has been already observed in~\cite{Alvarez:2014uda,delPino:2015mna} and in black hole solutions~\cite{Garcia:2003nm,Obukhov:2003sm} of the Mielke--Baekler model~\cite{Mielke:1991nn}. Notice that the Riemannian AdS curvature radius do not coincide with the one associated to the Riemannian--Cartan geometry denoted by $\ell_i$. This can be seen from the decomposition $\omega^{ab} = \tilde{\omega}^{ab} + K^{ab}$, where $\tilde{\omega}^{ab}$ is the Levi-Civita connection satisfying $\diff{e^a} + \tilde{\omega}^{a}{}_b\wedge e^b=0$ and $K^{ab}$ is the contorsion $1$-form defined as $T^a = K^{a}{}_b\wedge e^b$. This decomposition yields to
\begin{align}
 R^{ab} = \tilde{R}^{ab} + \tilde{\Diff{}}K^{ab} + K^{a}{}_c\wedge K^{cb},
\end{align}
where $\tilde{R}^{ab} = \diff{\tilde{\omega}^{ab}} + \tilde{\omega}^{a}{}_c\wedge\tilde{\omega}^{cb}$ and $\tilde{\Diff{}}$ is the Lorentz-covariant derivative with respect to $\tilde{\omega}^{ab}$. Thus, the Riemannian curvature $2$-forms of the $3$-dimensional section are locally constant and given by
\begin{align}
 \tilde{\bar{R}}_{(i)}^{\bar{a}\bar{b}} = -\frac{1}{\tilde{\ell}_i^2} e^{\bar{a}}\wedge e^{\bar{b}}.
\end{align} 
Interestingly, the case with scalar-Gauss--Bonnet coupling admits a Riemannian flat geometry when the condition $2\kappa - \alpha_3^2\Lambda^2 = 0$ is satisfied, even though the negative bare cosmological constant is nonvanishing. The case when $2\kappa - \alpha_3^2\Lambda^2<0$ and $\Lambda<0$ admits a positive curvature radius, however, it represents a naked singularity. The torsional invariants reported in Refs.~\cite{Sezgin:1981xs,Diakonov:2011fs,Baekler:2011jt} have been computed in all cases and it turns out that they are locally constant everywhere. However, the nature of the singularity at $r=0$ persists according to the BTZ geometry~\cite{Banados:1992gq1,Banados:1992gq2}. 

It is well-known that Dirac spinors are sensitive to the axial piece of torsion~\cite{Hehl:1976kj,Blagojevic:2013xpa}, defined in terms of the irreducible components\footnote{In fact, Dirac fermions are source of the axial component of torsion when backreaction is considered.} 
\begin{align}\label{conirred}
K^{ab} &= V^{[a}e^{b]} + \epsilon^{ab}{}_{cd}A^c e^d + Q^{ab},
\end{align}
where $V^a$ and $A^a$ are Lorentz-valued $0$-forms denoting the vectorial and axial pieces, respectively, while the mixed part $Q^{ab}=Q^{ab}{}_c\,e^c$ satisfies $i_a Q^{ab} = 0 = Q_{ab}\,e^a\wedge e^b$, where $i_a$ is the inner contraction along the vector basis $E_a = E^{\mu}{}_a\partial_\mu$, such that $e^{a}{}_\mu E^{\mu}{}_b = \delta^a_b$ and $e^{a}{}_\mu E^{\nu}{}_a = \delta^\mu_\nu$. The Dirac equation in Riemann--Cartan spacetimes is given by
\begin{align}\label{diracforms}
 \star\gamma\wedge\left(\Diff{\psi} - \frac{1}{2} i_a T^a \psi \right) &= 0,
\end{align}
where $\gamma= \gamma_a e^a$ is the gamma matrix $1$-form satisfying the Clifford algebra $\left\{\gamma_a,\gamma_b \right\}=2\eta_{ab}$, and $\Diff{\psi} = \diff{\psi} + \tfrac{1}{4}\omega^{ab}\gamma_{ab}$ with $\gamma_{a_1...a_p} \equiv \gamma_{[a_1}...\gamma_{a_p]}$.
It is found that the solutions with scalar-Nieh--Yan and scalar-Pontryagin couplings possess nonvanishing axial torsion, thus, Dirac fermions will be sensitive to such black string backgrounds. For scalar-Gauss--Bonnet coupling, however, only the vectorial irreducible component of the torsion is present, therefore, fermions will \emph{not} be sensitive to the torsional part of this black string configuration. 

Additionally, it is well-known that point particles without spin move along geodesics, regardless the dynamical content of the gravitational theory. In Ref.~\cite{HEHL1971225}, an equation for the trajectory of spinning test particles was derived, showing that they are sensitive to all the components of the contorsion. This kind of test particles can serve to probe all the black string geometries with nonvanishing torsion presented in this work.

\section{Conclusions\label{sec:conclusions}}

In this work, different static and rotating four-dimensional black string solutions have been presented in vacuum within dynamical CSMG, with and without torsion. These solutions represent the first static and rotating black strings in dynamical CSMG with nontrivial contributions of the Cotton tensor. They represent the black string extension of the rotating BTZ black hole~\cite{Banados:1992wn} with one additional extended direction. For a horizon to exist, the same conditions of the BTZ black hole must hold. The solutions differ from the one reported in~\cite{Ertem:2009ur}, since the scalar field is dynamical and it can be completely determined from the field equations. Moreover, these configurations belong to the Chern--Simons sector of the space of solutions according to~\cite{Grumiller:2007rv}.

Next, the first-order formulation of CSMG is studied by considering nonminimal couplings of the scalar fields to the Nieh--Yan, Pontryagin, and Gauss--Bonnet densities. Homogeneous and rotating BTZ black strings are found that, to the best of the authors' knowledge, represent the first black strings with nonvanishing torsion reported in the literature.\footnote{In the context of string theory, a black string configuration supported by a completely antisymmetric 3-form was obtained in~\cite{Ginsparg:1992af}. This notion of ``torsion'' is not the Cartan's torsion, which is described by a 2-form in any spacetime dimension (see Ref.~\cite{Blagojevic:2013xpa}).} It is found that either axial or vectorial components of the torsion arise, while the other components are zero in all cases. We found that torsion contributes to an effective cosmological constant, shifting the Riemannian AdS curvature radius. This behaviour could have interesting consequences in cosmology by considering a torsion-driven acceleration without the cosmological constant and its fine-tuning problem. On the other hand, in the case of the scalar-Gauss--Bonnet coupling, there exist a particular choice of the coefficients that allows for a flat Riemannian geometry in presence of a nonvanishing cosmological constant. This is possible due to the presence of a vectorial component of the torsion that can cancel the contribution of the bare cosmological constant. Finally, it is shown that spinning test particles and Dirac spinors will sensitive to some of the BTZ black string backgrounds reported here, providing a suitable scenario for the latter to study spinorial quasinormal modes and their stability.

Charged black strings in dynamical CSMG were sought in this work, nevertheless, the same impossibility arising in the black string extension of the Reissner--Norstrom black hole was found (see~\cite{Giacomini:2018sho} and references therein). It is worth mentioning that this can be circumvented in higher dimensional gravity by considering Einstein--Gauss--Bonnet theory, $p$-forms instead of Maxwell fields~\cite{Giacomini:2018sho}, or in four-dimensional Einstein-$SU(2)$ Skyrme model where the charge like term comes from the inclusion of the Skyrme fields rather than from the Maxwell fields~\cite{Astorino:2018dtr}.

Interesting questions remain open. For instance, conserved Noether charges within the first-order formalism and their connection with black hole thermodynamics have been studied in~\cite{Nester:1991yd,Chen:1994qg,Chen:1998aw,Yo:1999ex,Yo:2001sy,Jacobson:2015uqa,Blagojevic:2002du,Blagojevic:2013xpa}. In order to apply these techniques for the black string solutions presented here, the gravitational degrees of freedom should be extended by considering the scalar fields present in CSMG. This is certainly of great interest and it is left for a future contribution. On the other hand, although the four-dimensional Schwarzschild black hole is stable under linear perturbations~\cite{Regge:1957td,Edelstein:1970sk,Wald1,Wald2}, it has been shown that its cylindrical extensions, alongside a variety of black strings and branes in $D\geq5$, suffer from the so-called Gregory--Laflame instability~\cite{Gregory:1993vy,Gregory:1994bj}. It remains to be seen if torsion modifies this instability somehow. If these black string are unstable, it is very interesting to figure out their final state from both Riemannian and non-Riemannian viewpoints, which could lead to the formation of naked singularities in four dimensions. 
%It is therefore interesting to address this issue in the future for the black string configurations presented in this work, and to study the role of torsional perturbations of the solutions.

\begin{acknowledgement}
The authors thank to Y.~Bonder, C.~Erices, D.~Hilditch, F.~Izaurieta, B.~Ju\'arez-Aubry, R.~Olea, J.~Oliva, H.~Quevedo, and M.~Salgado for valuable comments and remarks. The authors also thank the anonymous referee for his/her thorough revision that yielded the manuscript into a neater form. A. C. is supported by Fondo Nacional de Desarrollo Cient\'ifico y Tecnol\'ogico Grant No. 11170274 and Proyecto Interno Ucen I+D-2016, CIP2016. C.C. is supported by UNAM-DGAPA-PAPIIT Grant RA101818 and UNAM-DGAPA postdoctoral fellowship. C.C. thanks to Universidad de Concepci\'on and Universidad Central de Chile for hospitality during early stages of this work.
\end{acknowledgement}

\appendix
\section{Decompositions}\label{decomp}
The curvature and torsion constructed out of Eqs.~\eqref{vierbeinansatz} and~\eqref{spinansatz} can be decomposed as
\begin{align}
 R^{ab} &= \begin{cases}
            R^{\bar{a}\bar{b}} &= M^{\bar{a}\bar{b}} + N^{\bar{a}\bar{b}} e^3,\\
            R^{\bar{a}3} &= L^{\bar{a}} + Q^{\bar{a}} e^3,
           \end{cases}\\           
 T^{a} &= \begin{cases}
           T^{\bar{a}} &= \bar{T}^{\bar{a}} + P^{\bar{a}} e^3,\\
           T^3 &= -\beta_{\bar{b}}e^{\bar{b}} + \gamma_{\bar{b}}e^{\bar{b}} e^3,
          \end{cases}          
\end{align}
where $\bar{R}^{\bar{a}\bar{b}} = \diff{\bar{\omega}^{\bar{a}\bar{b}} + \bar{\omega}^{\bar{a}}}{}_{\bar{c}}\wedge \bar{\omega}^{\bar{c}\bar{b}}$ and $\bar{T}^{\bar{a}} = \diff{\bar{e}^{\bar{a}}} + \bar{\omega}^{\bar{a}}{}_{\bar{b}}\wedge \bar{e}^{\bar{b}}$, with
\begin{align}
 M^{\bar{a}\bar{b}} &= \bar{R}^{\bar{a}\bar{b}} - \beta^{\bar{a}}\wedge\beta^{\bar{b}},\\
 N^{\bar{a}\bar{b}} &= \bar{\Diff{}}\alpha^{\bar{a}\bar{b}} - 2\beta^{[\bar{a}}\gamma^{\bar{b}]},\\
 L^{\bar{a}} &= \bar{\Diff{}}\beta^{\bar{a}},\\
 Q^{\bar{a}} &= \bar{\Diff{}}\gamma^{\bar{a}} - \alpha^{\bar{a}}{}_{\bar{b}}\beta^{\bar{b}},\\
 P^{\bar{a}} &= \beta^{\bar{a}} - \alpha^{\bar{a}}{}_{\bar{b}} e^b,
\end{align}
where $\bar{\Diff{}} = \diff{} + \bar{\omega}$. The energy-momentum $3$-form~\eqref{tauscal} can be written as
\begin{align}
 \tau^{(i)}_a &= \begin{cases}
            \tau^{(i)}_{\bar{a}} &= \bar{\tau}^{(i)}_{\bar{a}} + \chi^{(i)}_{\bar{a}}\wedge e^3 ,\\
            \tau^{(i)}_3 &= \mu^{(i)} + \nu^{(i)}\wedge e^3,
           \end{cases}
\end{align}
where
\begin{align}
 \bar{\tau}^{(i)}_{\bar{a}} &= \frac{1}{2}\left[\frac{1}{2}\partial_{\bar{b}}\varphi_i\partial^3\varphi_i \epsilon_{\bar{a}\bar{l}\bar{m}}e^{\bar{b}}\wedge e^{\bar{l}}\wedge e^{\bar{m}} + \frac{1}{3!}\partial_{\bar{a}}\varphi_i \partial^3 \varphi_i\epsilon_{\bar{c}\bar{d}\bar{e}} e^{\bar{c}}\wedge e^{\bar{d}}\wedge e^{\bar{e}}  \right],\\
 \chi^{(i)}_{\bar{a}} &= \frac{1}{2}\left[\partial_{\bar{b}}\varphi_i \partial^{\bar{c}} \varphi_i \epsilon_{\bar{a}\bar{c}\bar{m}} e^{\bar{b}}\wedge e^{\bar{m}} + \frac{1}{2}\left(\partial_z \varphi_i\right)^2\epsilon_{\bar{a}\bar{l}\bar{m}}e^{\bar{l}}\wedge e^{\bar{m}} - \frac{1}{2}\partial_{\bar{a}}\varphi_i\partial^{\bar{b}}\varphi_i \epsilon_{\bar{b}\bar{d}\bar{e}} e^{\bar{d}}\wedge e^{\bar{e}} \right],\\
 \mu^{(i)} &= -\frac{1}{2}\left[\frac{1}{2}\partial_{\bar{b}}\varphi_i \partial^{\bar{c}}\varphi_i\epsilon_{\bar{c}\bar{l}\bar{m}}e^{\bar{b}}\wedge e^{\bar{l}}\wedge e^{\bar{m}} - \frac{1}{3!}\left(\partial_z\varphi_i\right)^2\epsilon_{\bar{c}\bar{d}\bar{e}}e^{\bar{c}}\wedge e^{\bar{d}}\wedge e^{\bar{e}} \right],\\
 \nu^{(i)} &= -\frac{1}{4}\partial_z\varphi_i \partial^{\bar{c}}\varphi_i\epsilon_{\bar{c}\bar{l}\bar{m}} e^{\bar{l}}\wedge e^{\bar{m}},
\end{align}
and $\epsilon_{\bar{a}\bar{b}\bar{c}3} \equiv \epsilon_{\bar{a}\bar{b}\bar{c}}$. 

Similarly, the field equations~\eqref{eomfocsmg} admit the decomposition
\begin{align}
 \mathcal{E}_a &= \begin{cases}
                   \mathcal{E}_{\bar{a}} &= \mathcal{A}_{\bar{a}} + \mathcal{B}_{\bar{a}} e^3, \\
                   \mathcal{E}_3 &= \mathcal{C} + \mathcal{D} e^3 
                  \end{cases}
\end{align}
where
\begin{align}\label{adef}
 \mathcal{A}_{\bar{a}} &= -2\epsilon_{\bar{a}\bar{b}\bar{c}} L^{\bar{b}}\wedge e^{\bar{c}} + \alpha_1 \partial_{\bar{b}}\varphi_1 e^{\bar{b}}\wedge \bar{T}_{\bar{a}} - 2\kappa\sum_{i=1}^{3}\bar{\tau}^{(i)}_{\bar{a}},\\
 \mathcal{B}_{\bar{a}} &= \epsilon_{\bar{a}\bar{b}\bar{c}}\left( M^{\bar{b}\bar{c}} - \Lambda e^{\bar{b}}\wedge e^{\bar{c}} + 2 Q^{\bar{b}}\wedge e^{\bar{c}}\right) + \alpha_1\left(\partial_{\bar{b}}\varphi_1 e^{\bar{b}}\wedge P_{\bar{a}} + \partial_z \varphi_1\bar{T}_{\bar{a}} \right) - 2\kappa\sum_{i=1}^{3}\chi^{(i)}_{\bar{a}} ,\\
 \mathcal{C} &= -\epsilon_{\bar{a}\bar{b}\bar{c}}\left( M^{\bar{a}\bar{b}} - \frac{\Lambda}{3} e^{\bar{a}}\wedge e^{\bar{b}} \right)\wedge e^{\bar{c}} - \alpha_1 \partial_{\bar{a}} \varphi_1 e^{\bar{a}}\wedge \beta_{\bar{b}}\wedge e^{\bar{b}} - 2\kappa\sum_{i=1}^{3}\mu^{(i)},\\
 \label{ddef}
 \mathcal{D} &= \epsilon_{\bar{a}\bar{b}\bar{c}}N^{\bar{a}\bar{b}}\wedge e^{\bar{c}} + \alpha_1\left(\partial_{\bar{a}}\varphi_1 e^{\bar{a}}\gamma_{\bar{b}} - \partial_z\varphi_1\beta_{\bar{b}} \right)\wedge e^{\bar{b}} - 2\kappa\sum_{i=1}^{3}\nu^{(i)},
\end{align}
and
\begin{align}
 \mathcal{E}_{ab} &= \begin{cases}
                      \mathcal{E}_{\bar{a}\bar{b}} &= \mathcal{W}_{\bar{a}\bar{b}} + \mathcal{X}_{\bar{a}\bar{b}} e^3,\\
                      \mathcal{E}_{\bar{a}3} &= \mathcal{Y}_{\bar{a}} + \mathcal{Z}_{\bar{a}} e^3
                     \end{cases}
\end{align}
where
\begin{align}\label{wdef}
 \mathcal{W}_{\bar{a}\bar{b}} &= \epsilon_{\bar{a}\bar{b}\bar{c}} \beta_{\bar{d}}\wedge e^{\bar{d}}\wedge e^{\bar{c}} - \frac{\alpha_1}{2}\partial_{\bar{c}}\varphi_1 e^{\bar{c}}\wedge e_{\bar{a}}\wedge e_{\bar{b}} - \alpha_2 \partial_{\bar{c}}\varphi_2 e^{\bar{c}}\wedge M_{\bar{a}\bar{b}} + 2\alpha_3 \epsilon_{\bar{a}\bar{b}\bar{c}}\partial_{\bar{d}}\varphi_3 e^{\bar{d}}\wedge L^{\bar{c}},\\
 \notag
 \mathcal{X}_{\bar{a}\bar{b}} &= \epsilon_{\bar{a}\bar{b}\bar{c}}\left(\bar{T}^{\bar{c}} + \gamma_{\bar{d}}e^{\bar{d}}\wedge e^{\bar{c}} \right) - \frac{\alpha_1}{2}\partial_z\varphi_1 e_{\bar{a}} \wedge e_{\bar{b}} - \alpha_2\left(\partial_{\bar{c}}\varphi_2 e^{\bar{c}}\wedge N_{\bar{a}\bar{b}} + \partial_z\varphi_2 M_{\bar{a}\bar{b}} \right) \\
 &\quad + 2\alpha_3 \epsilon_{\bar{a}\bar{b}\bar{c}}\left(\partial_{\bar{d}}\varphi_3 e^{\bar{d}}\wedge Q^{\bar{c}} + \partial_z \varphi_3 L^{\bar{c}} \right),\\
 \label{ydef}
 \mathcal{Y}_{\bar{a}} &= \epsilon_{\bar{a}\bar{b}\bar{c}}\bar{T}^{\bar{b}}\wedge e^{\bar{c}} - \alpha_2 \partial_{\bar{b}}\varphi_2 e^{\bar{b}} \wedge L_{\bar{a}} + \alpha_3 \epsilon_{\bar{a}\bar{b}\bar{c}} \partial_{\bar{d}}\varphi_3 e^{\bar{d}}\wedge M^{\bar{b}\bar{c}},\\
 \notag
 \mathcal{Z}_{\bar{a}} &= -\epsilon_{\bar{a}\bar{b}\bar{c}}P^{\bar{b}}\wedge e^{\bar{c}} - \frac{\alpha_1}{2}\partial_{\bar{b}}\varphi_1 e^{\bar{b}}\wedge e_{\bar{a}} - \alpha_2 \left(\partial_{\bar{b}}\varphi_2 e^{\bar{b}}\wedge Q_{\bar{a}} + \partial_z\varphi_2 L_{\bar{a}} \right) \\
 &\quad + \alpha_3\epsilon_{\bar{a}\bar{b}\bar{c}}\left(\partial_{\bar{d}}\varphi_3 e^{\bar{d}}\wedge N^{\bar{b}\bar{c}} + \partial_z\varphi_3 M^{\bar{b}\bar{c}} \right).
\end{align}
On shell, each $\mathcal{A}_{\bar{a}}$, $\mathcal{B}_{\bar{a}}$, $\mathcal{C}$, $\mathcal{D}$, $\mathcal{W}_{\bar{a}\bar{b}}$, $\mathcal{X}_{\bar{a}\bar{b}}$, $\mathcal{Y}_{\bar{a}}$, and $\mathcal{Z}_{\bar{a}}$ vanish independently.

\section{Riemannian Chern--Simons modified gravity in the language of diferential forms\label{sec:relation}}

In this Appendix, it is shown that the BTZ black strings solution of Sec.~\ref{sec:rbs} can be obtained from the first-order formalism, by imposing the torsion-free condition through a Lagrange multiplier. The action under consideration is given by 
\begin{align}\notag
 S[e^a,\omega^{ab},\varphi_i,\zeta_a] &= \frac{1}{4\kappa}\int\left[\epsilon_{abcd}\left(R^{ab} - \frac{\Lambda}{6}e^a\wedge e^b\right)\wedge e^c\wedge e^d + \alpha\varphi R^{a}{}_b\wedge R^{b}{}_a + \zeta_a\wedge T^a \right] \\ 
 \label{focsmgalag}
 &\quad - \frac{1}{2}\int\diff{\varphi}\wedge\star\diff{\varphi},
\end{align}
where $\zeta_a$ is a Lorentz-valued Lagrange multiplier $2$-form. The field equations for the vierbein, Lorentz connection, scalar fields, and Lagrange multiplier are respectively given by
\begin{subequations}\label{eomfocsmgl}
\begin{align}\label{eomel}
 \mathcal{E}_a + \frac{1}{2}\Diff{\zeta_a} &= 0,\\
 \label{eomwl}
 \mathcal{E}_{ab} + \frac{1}{2}\zeta_{[a}\wedge e_{b]} &= 0,\\
 \mathcal{E}^{(\varphi_i)} &= 0,\\
 T^a &= 0,
\end{align}
\end{subequations}
where $\mathcal{E}_a$, $\mathcal{E}_{ab}$, and $\mathcal{E}^{(\varphi_i)}$ have been defined in Eq.~\eqref{eomfocsmg}, by taking $\varphi_1 = \varphi_3 = 0$ and redefining $\varphi_2 \equiv \varphi$. The Lagrange multiplier can be solved algebraically from Eq.~\eqref{eomwl} as
\begin{align}
 \zeta_a = 4 i^b \mathcal{E}_{ba} - \left(i^b i^c \mathcal{E}_{bc} \right) e_a.
\end{align}
Using this expression, alongside the constraint $T^a = 0$, the field equations reduces to
\begin{subequations}\label{eomrl}
\begin{align}
 \epsilon_{abcd}\left(\tilde{R}^{bc} - \frac{\Lambda}{3} e^b\wedge e^c\right)\wedge e^d + \frac{1}{2}\tilde{\Diff{}}\zeta_a - 2\kappa \tau_a^{(\varphi)} &= 0,\\
 \diff{\star}\diff{\varphi} + \frac{\alpha}{4\kappa} \tilde{R}^{a}{}_b\wedge\tilde{R}^{b}{}_a &= 0,
\end{align}
 \end{subequations}
where, as before, tilde denote Riemannian (torsion-free) quantities. It is straightforward to see that Eq.~\eqref{solstatic}, as well as Eq.~\eqref{rota} with~\eqref{solrota}, solve the field equations~\eqref{eomrl}.

\bibliographystyle{spphys}
\bibliography{BS}

\end{document}